\documentclass[conference]{IEEEtran}

\IEEEoverridecommandlockouts
\usepackage{cite}
\usepackage{amsmath,amssymb,amsfonts}
\usepackage{algorithmic}
\usepackage{graphicx}
\usepackage{textcomp}
\usepackage{xcolor}
\usepackage{diagbox}

\usepackage{tikz}
\usetikzlibrary{shapes,arrows}
\usetikzlibrary{calc}
\usepackage{amsmath,bm,times}

\def\BibTeX{{\rm B\kern-.05em{\sc i\kern-.025em b}\kern-.08em
    T\kern-.1667em\lower.7ex\hbox{E}\kern-.125emX}}

\renewcommand{\baselinestretch}{0.92}

\begin{document}

\title{Physics-Informed Neural Networks \\ for Power Systems 
}

\author{\IEEEauthorblockN{George S. Misyris, Andreas Venzke and Spyros Chatzivasileiadis \vspace{-8pt}}\\
\IEEEauthorblockA{Center for Electric Power and Energy\\
Technical University of Denmark\\
\{gmisy, andven, spchatz\}@elektro.dtu.dk}
}

\maketitle
\begin{abstract}
This paper introduces for the first time, to our knowledge, a framework for physics-informed neural networks in power system applications. Exploiting the underlying physical laws governing power systems, and inspired by recent developments in the field of machine learning, this paper proposes a neural network training procedure that can make use of the wide range of mathematical models describing power system behavior, both in steady-state and in dynamics. Physics-informed neural networks require substantially less training data and can result in simpler neural network structures, while achieving high accuracy. This work unlocks a range of opportunities in power systems, being able to determine dynamic states, such as rotor angles and frequency, and uncertain parameters such as inertia and damping at a fraction of the computational time required by conventional methods. This paper focuses on introducing the framework and showcases its potential using a single-machine infinite bus system as a guiding example. Physics-informed neural networks are shown to accurately determine rotor angle and frequency up to \emph{87 times faster} than conventional methods.
\end{abstract}


\vspace{-3pt}
\section{Introduction}
\vspace{-1pt}
Machine learning techniques demonstrate impressive results for a range of highly complex tasks, especially where an accurate mathematical representation of the problem cannot be obtained. Applications include image recognition, robotics, weather forecasting, and others \cite{lecun2015deep}. In power
systems, decision trees and neural networks have been shown to solve computational problems both in dynamics and optimization at a fraction of the time required by traditional approaches, being up to three order of magnitude faster \cite{wehenkel2012automatic,donnot2017introducing, sun2018deep, arteaga2019deep, ferdin2019predicting}.

Up to this point, however, machine learning methods applied to power systems (and other physical systems) were largely agnostic to the underlying physical model. This made them heavily dependent on the quality of the training data, it required large training datasets, and oftentimes complex neural network structures. Despite recent efforts for efficient creation of datasets with encouraging results \cite{thams2019efficient, venzke2019efficient}, generating the required training dataset size still requires substantial computational effort. In this work, inspired by \cite{chen2018neural,RAISSI2019686}, we reduce the dependency on training data and complex neural network structures by exploiting \emph{inside the neural network training} the underlying physical laws described by power system models. 

This is the first work, to our knowledge, that proposes physics-informed neural networks for power system applications. It introduces a neural network training framework that can exploit the underlying physical laws and the available power system models both for steady-state and dynamics. Following recent approaches reported in \cite{chen2018neural,RAISSI2019686}, we incorporate the power system differential and algebraic equations inside the training procedure. Exploiting advances in automatic differentiation \cite{baydin2018automatic} that are implemented in Tensorflow \cite{abadi2016tensorflow}, we can directly compute derivatives of neural network outputs during training, such as the rotor angle, and build neural networks able to accurately capture the rotor angle and frequency dynamics. Our approach (i) requires less initial training data, (ii) can result to smaller neural networks, while (iii) demonstrating high performance. 

Physics-informed neural networks introduce a novel technology that may lead to a new class of numerical solvers \cite{RAISSI2019686} as well as dynamic state estimation techniques \cite{Zhao2019PSDSE}. Within power systems, they have the potential to solve systems of differential-algebraic equations at a fraction of computational time required for conventional methods, are able to directly determine the value of state variables at any time instant $t_1$ (without the need to integrate from $t_0$ to $t_1$), and can handle directly higher-order differential equations without the need to introduce additional variables to solve a first-order system. 


In this paper, we present the main principles for the application of physics-informed neural networks of \cite{RAISSI2019686} in power systems, focusing on power system dynamics and using the swing equation as an example. Besides obtaining solutions to ordinary differential equations, we demonstrate how the same methods can be used to estimate uncertain parameters such as inertia and damping. The contributions of our work are:
\begin{enumerate}
    \item We propose physics-informed neural networks to (i) accurately determine solutions of differential equations and, thus, values of power system dynamic states, such as rotor angle and frequency, and (ii) identify uncertain power system parameters. Contrary to previous approaches, physics-informed neural networks utilize the underlying physical model, lead to significantly reduced computation time and need less training data. 
    \item  For the single machine infinite bus (SMIB) system, we show that physics-informed neural networks (i)~predict system dynamics with high accuracy at a fraction of the computational time required by conventional approaches (28-87 times faster in our study), and (ii)~can identify with high accuracy uncertain system parameters such as inertia and damping.
\end{enumerate} 

This paper is structured as follows: Section~\ref{Meth} describes the employed power system model and introduces the architecture of physics-informed neural networks. Section III presents simulation results demonstrating the performance of physics-informed neural networks. Section~IV discusses the challenges and the opportunities emerging from the successful application of this concept. Section~V concludes. The code to reproduce the simulation results is available online \cite{GitHubDatabase}.

\section{Methodology} \label{Meth}
\subsection{Physical Model for Power System Dynamics} 
Power system dynamics, in their simplest and most common form, are described by the swing equation, neglecting transmission losses and bus voltage deviations. For each generator $k$, the resulting system of equations can then be represented by \cite{PESGM_2016A,Misyris2018}:
\begin{equation}
\centering
    m_k\ddot{\delta_k}+d_k\dot{\delta_k}+\sum_jB_{kj}V_kV_j\sin(\delta_k-\delta_j)-P_k=0
    \label{Eq::SwingEquation}
\end{equation}
where $m_k$ defines the generator inertia constant, $d_k$ represents the damping coefficient, $B_{kj}$ is the $\{k,j\}\textnormal{-entry}$ of the bus susceptance matrix, $P_k$ is the mechanical power of the $k^{\rm th}$ generator, $V_k, V_j$ are the voltage magnitudes at buses $k,j$ and $\delta_k, \delta_j$ represent the voltage angles behind the transient reactance. $\dot\delta_k$ is the angular frequency of generator $k$, often also denoted as $\omega_k$. 
\subsubsection{Single Machine Infinite Bus (SMIB) System} 
\begin{figure}
    \centering
    \includegraphics[width=0.8\linewidth, trim={0cm 0.0cm 0cm 0.5cm}, clip]{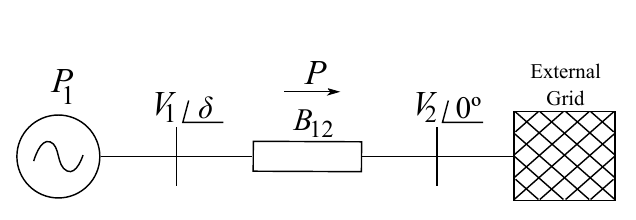}
    \vspace{-10pt}
    \caption{Single machine infinite bus system}
    \label{fig:SMIB}
    \vspace{-15pt}
\end{figure}

The single-machine infinite-bus system, shown in Fig.~\ref{fig:SMIB}, has been widely used to understand and analyze the fundamental dynamic phenomena occurring in power systems. As the focus of this paper is on the introduction of physics-informed neural networks for power systems, we will use this system as a guiding example. Note though that our proposed framework is general. Future work will focus on larger, more complex systems.
The swing equation \eqref{Eq::SwingEquation} for the SMIB system is given by:
\begin{align}
        m_1\ddot{\delta}+d_1\dot{\delta}+B_{12}V_1V_2\sin(\delta)-P_1=0 \label{Eq:Swing_SMIB}
\end{align}
In the rest of this paper, we will show how physics-informed neural networks can accurately estimate both rotor angle $\delta$ and frequency $\dot{\delta}$ while $P_1$ varies within $[P_{\rm min}, P_{\rm max}]$, and can identify uncertain parameters such as $m_1$ and $d_1$.
\subsection{Physics-Informed Neural Networks}
In the following, we explain the general architecture of physics-informed neural networks, and detail its application to the SMIB system. Feed-forward neural networks are composed of the input layer, fully connected hidden layers having a non-linear activation function at each neuron, and the output layer. Between each layer a weight matrix $\mathbf{W}$ and bias $\mathbf{b}$ is applied. During training, weight matrices and biases are optimized to minimize an objective function which usually penalizes the deviation of the neural network prediction from the training data. Neural networks are universal function approximators as they can, in theory, learn any unknown function between some inputs and outputs. Therefore, neural networks could be used to directly learn the nonlinear mapping between the inputs and the outputs of differential equations, such as \eqref{Eq:Swing_SMIB}. Not taking into account the underlying physical model, however, will require large amounts of training data and a large neural network size. The work in \cite{RAISSI2019686} introduced a framework for physics-informed neural networks which we will rely on in the following. Considering physical laws during training allows to bound the space of admissible solutions to the neural network parameters, which translates to a lower requirement in both the amount of training data and neural network size. 

Following notation similar to Ref.~\cite{RAISSI2019686}, the general form of the functions that the physics-informed neural network can approximate is:
\begin{equation}
    \centering
    \frac{\partial u}{\partial t} = - N[u;\lambda] \>, \>  x \in \Omega, \> t \in [0, T]
    \label{Eq::GeneralForm}
\end{equation}
where $u(t,x)$ is the solution and $N[u; \lambda]$ is a nonlinear operator connecting the state variables $u$ with the system parameters $\lambda$. The term $t$ denotes time and $x$ the system input. The domain $\Omega$ can be bounded based on prior knowledge of the dynamical system and $[0, T]$ is the time interval within which the system evolves. The model parameters $\lambda$ can be constant or unknown. In case $\lambda$ is unknown, the problem of approximating function \eqref{Eq::GeneralForm} becomes a problem of system identification, where we seek parameters $\lambda$ for which the expression in \eqref{Eq::GeneralForm} is satisfied. 
\begin{figure}[t]
    \centering
\resizebox{0.45\textwidth}{!}{
\begin{tikzpicture}[>=stealth]
  \coordinate (orig)   at (0,0);
  \coordinate (LLD)    at (4,0);
  \coordinate (AroneA) at (-1/2,11/2);
  \coordinate (ArtwoA) at (-1/2,5);
  \coordinate (ArthrA) at (-1/2,9/2);
  \coordinate (LLA)    at (1,4);
  \coordinate (LLB)    at (4,4);
  \coordinate (LLC)    at (7,4);
  \coordinate (AroneC) at (25/2,11/2);
  \coordinate (ArtwoC) at (25/2,5);
  \coordinate (ArthrC) at (25/2,9/2);
  \coordinate (conCBD) at (21/2,9/2);
  \coordinate (conCB)  at (21/2,7/2);
  \coordinate (coCBD)  at (11,5);
  \coordinate (coCB)   at (11,3);
  \coordinate (conCBA) at (23/2,11/2);
  \coordinate (conCA)  at (23/2,5/2);

  \node[draw, minimum width=2cm, minimum height=2cm, anchor=south west, text width=2cm, align=center, rounded corners] (A) at (LLA) {Neural\\Network \\ (NN)};
  \node[right of=(A)] (B) at (LLB) {};
  \draw[->] (AroneA) -- node[above]{$x$} ($(A.180) + (0,1/2)$);
  \draw[->] (ArthrA) -- node[above]{$t$} ($(A.180) + (0,-1/2)$);

  \draw[->] ($(A.0)+(0,1/2)$) -- node[above] {$u(t,x)$} ($(B.180) + (3.35,1.5)$);
  \path[fill] ($(A.0)+(0.5,1/2)$) circle[radius=2pt] ; 
   \draw ($(A.0)+(0.5,1/2)$) -- ($(A.0)+(0.5,-1/2)$);
    \draw[->] ($(A.0)+(0.5,-1/2)$)-- ($(A.0)+(1,-1/2)$);
  \node[draw, minimum width=2.5cm, minimum height=1cm, anchor=south west, text width=2.5cm, align=center, rounded corners] (BB) at ($(A.0)+(1,-1.25)$) {{\footnotesize Differentiate}\\ {\footnotesize NN Output $u(t,x)$} \\ {\footnotesize and apply \eqref{eq:f_NN}}};
   \draw[->] ($(BB.0)+(0,0)$) -- node[above] {$f(t,x)$} ($(BB.0)+(1.25,0)$);
     
  








\end{tikzpicture}
}
    \caption{General structure of a physics-informed neural network: it predicts the output $u(t,x)$ given inputs $x$ and $t$. Then, using automatic differentiation \cite{baydin2018automatic} of the same neural network, the partial derivatives of $u(t,x)$ are computed, and $f(t,x)$ is evaluated. The parameters $\lambda$ are either assumed to be known, or are optimized as part of the neural network training. During training, the neural network weights and biases are adjusted according to loss function \eqref{Eq:meansquarederror}, which minimizes the deviation of both the output prediction $u(t,x)$ from ground truth and $f(t,x)$ from $0$.}
    \label{fig:Gen_NN}
    \vspace{-15pt}
\end{figure}
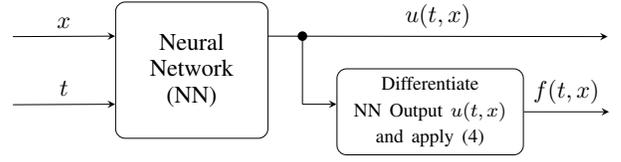{}
To enforce the physical law describing the dynamical system we define the physics-informed neural network $f(t,x)$:
\begin{equation}
    f(t,x) = \frac{\partial u}{\partial t} + N[u,\lambda] \label{eq:f_NN}
\end{equation}
Note that if the system parameters $\lambda$ are known the nonlinear operator $N[u,\lambda]$ simplifies to $N[u]$. The overall architecture is shown in Fig.~\ref{fig:Gen_NN}. A neural network is used to predict $u(t,x)$ based on the inputs $t$ and $x$. To determine $f(t,x)$, we use automatic differentiation \cite{baydin2018automatic} of the components of the neural network predicting $u(t,x)$. Based on this, we compute the required derivatives of $u(t,x)$ with respect to time $t$ and system inputs $x$. As a result, the neural network predicting $f(t,x)$ has the same parameters  compared to the neural network predicting $u(t,x)$, but different activation functions. The shared parameters of the two neural networks are optimized by minimizing the loss function:
\begin{equation}
    MSE = \underbrace{\tfrac{1}{N_u}\sum_i^{N_u}|u(t_u^i,x_u^i)-u^i|^2}_{MSE_u}+\underbrace{\tfrac{1}{N_f}\sum_i^{N_f}|f(t_f^i,x_f^i)|^2}_{MSE_f}
    \label{Eq:meansquarederror}
\end{equation}
where $MSE_u$ denotes the mean squared error loss corresponding to the initial data, $N_u$ is the total number of training data, $MSE_f$ is the mean squared error at a finite set of collocation points and $N_f$ is the total number of collocation points. The number of collocation points and training data influence the prediction accuracy and the computational time to optimize the loss function. The error $MSE_u$ enforces the boundary conditions of the independent variables $x$ and $MSE_f$ enforces the physics of the dynamical system imposed by the condition \eqref{Eq::GeneralForm}, i.e. it penalizes deviations of the predicted physical law. Given a training data set and known system parameters $\lambda$, we seek to find the parameters (weights and biases) of the neural networks which minimize \eqref{Eq:meansquarederror}. If the parameters $\lambda$ are unknown, we train for the same objective but consider the system parameters as additional variables. 

\subsubsection{Physics-informed neural networks capturing power system dynamics}
We show how physics-informed neural networks can be used to derive $\delta$ and $\omega=\dot{\delta}$ of the swing equation \eqref{Eq:Swing_SMIB} at any time instant $t$ and for a range of mechanical power $P_1$. We assume that the system parameters $\lambda:=\{m_1, d_1, B_{12}\}$ are known and the voltages $V_1$ and $V_2$ are fixed. As a result the system input is defined as $x:=\{P_1\}$. 
In contrast with conventional numerical solvers, which require the conversion of higher-order ordinary differential equations (ODEs) to first-order in order to solve them (by introducing additional variables), physics-informed neural networks can directly incorporate higher-order ODEs, as we show in \eqref{Eq:Fdelta}. 

Incorporating \eqref{Eq:Swing_SMIB} to the neural network, function \eqref{eq:f_NN} is given by:
\begin{align}
\centering
    u(t,x)&:=\delta(t, P_1), \\
    f_{\delta}(t,P_1) & = m_1\ddot{\delta} + d_1\dot{\delta} + B_{12}V_1V_2\sin(\delta) - P_1, \nonumber\\
    &\>\>\>\> P_1 \in [P_{\rm min}, P_{\rm max}],\>\> \>\> t\in[0, T]
    \label{Eq:Fdelta}
\end{align}
The interval $[0, T]$ can be defined based on the time period of interest for the dynamic simulation. The domain $\Omega$ of the input $P_1$ is restricted to $[P_{\rm min}, P_{\rm max}]$. The neural network output is $\delta(t, P_1)$. After the training phase, the frequency signal $\omega:= \dot{\delta}$ is extracted as a function of the estimated angle $\delta$. As a result, the prediction error of the frequency $\omega$ depends on the prediction error of the angle $\delta$ and the differential method.  
In the rest of the paper, we refer to this neural network structure as ${\rm NN}_{\delta}$. 

\subsubsection{Data-driven discovery of inertia and damping coefficients}
Information about power system parameters such as system inertia is of significant importance for system operators to prevent large frequency deviations and maintain frequency stability. As described in \cite{Misyris2018}, due to varying generation of converter-connected renewable energy sources, the inertia level of power systems becomes uncertain and has to be estimated (or predicted) at regular time intervals  \cite{Zografos2017}. 
Physics-informed neural networks can be used to address the problem of system identification and data-driven discovery of partial differential equations. For this case, we define $m_1$ and $d_1$ as unknown parameters in \eqref{Eq:Fdelta}. The structure of the physics-informed neural network remains the same, with the only difference that a subset of the system parameters $\lambda$ are now treated as additional variables when minimizing \eqref{Eq:meansquarederror} during neural network training. 

\section{Simulation \& Results} \label{SimRes}
\subsection{Simulation Setup} 
Besides an initial training set, to assess the neural network performance we also need an extensive test data set. To create the training and test data sets we use the numerical solver $ode45$ in MATLAB with a time step of $0.1s$ and time interval $T=[0, 20s]$, resulting in 201 time steps for each trajectory. The voltage magnitudes $V_{1}$ and $V_2$ are equal to 1 p.u. and $B_{12}$ = 0.2 p.u. In our first case study, we assume system inertia and damping are known, and that the system is not at an equilibrium. Assuming an uncertain active power input in the range $P_1=[0.08, 0.18]$ and initial values for $\delta$ and $\omega$ equal to \mbox{0.1 rad} and \mbox{0.1 rad/s}, we generate 100 trajectories. As a result, our entire test and training dataset consists of $20'100$ samples.  We consider the interval from $[0.08, 0.18]$ to show the capability of the physics-informed neural network to accurately predict trajectories for uncertain power injections. For values larger than 0.18, the system becomes unstable, and for values lower than 0.08 multiple oscillations occur. For these regimes, we observe lower prediction accuracy, and different trained physics-informed neural networks could be used to achieve high accuracy in each of these regimes.

In our second case study, inertia and damping are also unknown parameters. 
Given scattered observed data about active power, frequency and angle measurements, our goal is to identify the parameters $m_1$ and $d_1$ of \eqref{Eq:Fdelta}, as well as to obtain the trajectory of $\delta$. Considering that the levels of inertia and damping vary, we assign 10 different values to $m_1$ and $d_1$ that lie within the range of $[0.1, 0.4]$ and $[0.05, 0.15]$, respectively. To this end, for each of the 10 pairs $\{m_1, d_1\}$ we generate 40 trajectories. 

Next, before starting the training procedure, as usual for neural networks, we need to determine an appropriate number of hidden layers and number of neurons per layer, the amount of training data $N_u$ and the number of collocation points $N_f$. We carried an extensive investigation of the appropriate values for each of those parameters, assessing the relative $L_2$ error between the predicted and the exact solution of $\delta(t,P_1)$ and $\omega(t,P_1)$ for a range of different configurations. In the case studies, we report results only for the most suitable configuration, which achieved the lowest $L_2$ error. Similar to \cite{RAISSI2019686}, as the required amount of training data $N_u$ is very small (only 40 data points), we use a gradient-based optimization algorithm to optimize the loss function $MSE = MSE_u + MSE_f$ in \eqref{Eq:meansquarederror}. We perform neural network training and testing in TensorFlow on a laptop (Intel Core i7 3.9 GHz, 32-GB RAM, single NVIDIA GeForce 940MX 2-GB). The hidden layers of the neural network use hyperbolic tangent activation functions. The code to reproduce the results is available online \cite{GitHubDatabase}.



\subsection{Data-driven solution of frequency dynamics through physics-informed neural networks}
\begin{figure}
    \centering
    \includegraphics[width=\linewidth]{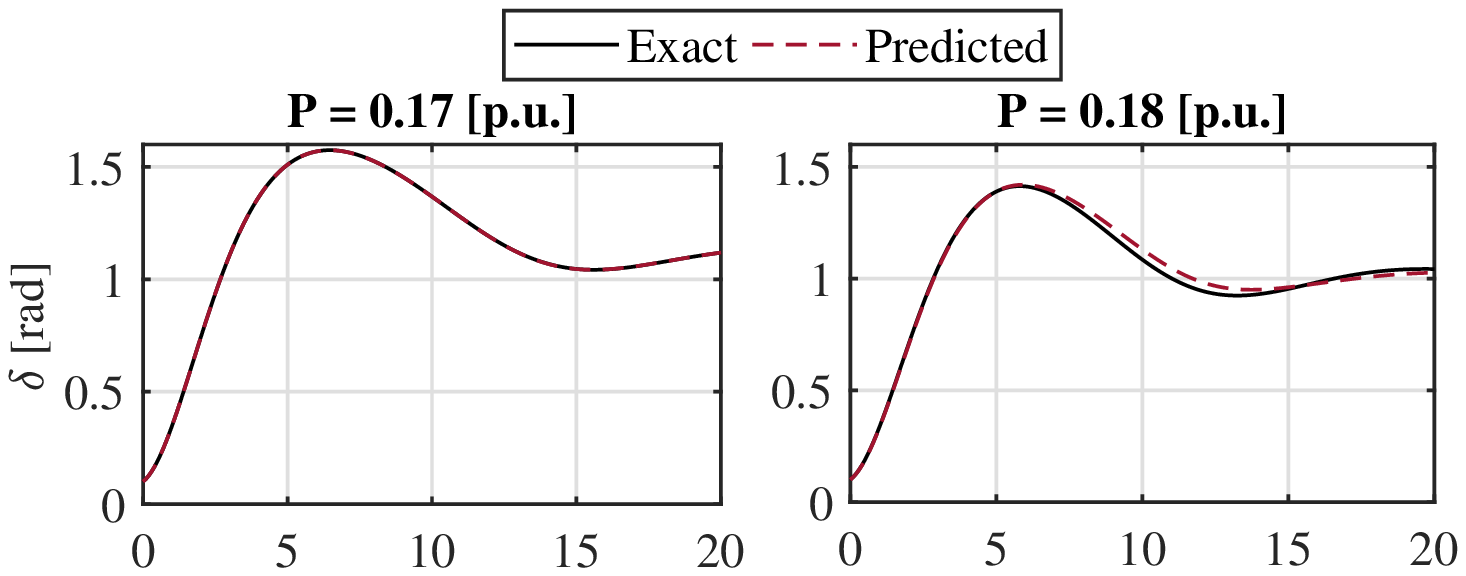}
    \includegraphics[width=\linewidth]{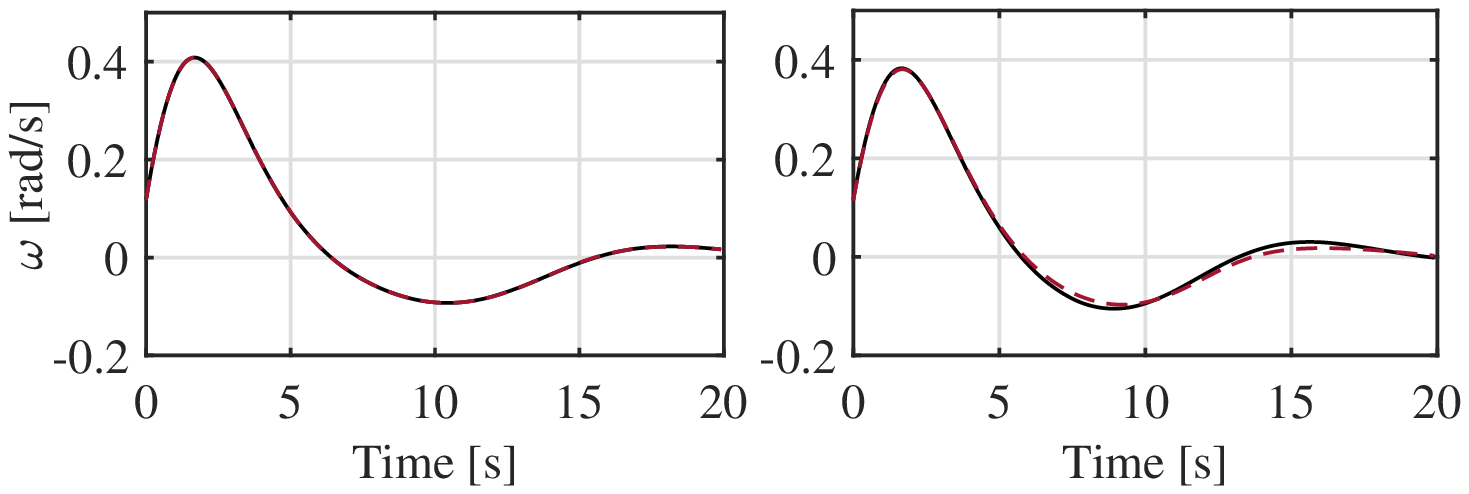}
    \vspace{-1cm}
    \caption{Comparison of the predicted and exact solution for the angle $\delta(t)$ and frequency $\omega(t)$ with the physics-informed neural network ${\rm NN}_{\delta}$. Note that to compute the frequency $\omega(t)$ we perform numerical differentiation of the angle $\delta(t)$ using a Newton method. In the left figures, we show the most accurate estimation of the trajectory of $\delta(t)$ and $\omega(t)$, with a relative $L_2$ error of $2.37\cdot10^{-2}$. In the right figures, we show the least accurate estimation of the trajectory of $\delta(t)$ and $\omega(t)$, with a relative $L_2$ error of $2.55\cdot10^{-4}$.} 
    \label{fig:DeltaAngle_OnlyAngle}
    \vspace{-15pt}
\end{figure}

The following parameters were selected to obtain the lowest $L_2$ error on the test data: we select a set of $N_u$ = 40 randomly distributed initial and boundary data across the entire spatio-temporal domain, \mbox{$N_f$ = $8'000$} collocation points, and a 5-layer neural network with 10 neurons per hidden layer. Observe that compared to conventional neural network approaches, we only need a very small amount of samples ($N_u$ = 40). Increasing $N_u$ in our simulations, led to over-fitting to the training data. 
Training took 223 seconds and the relative $L_2$ error between exact and predicted solutions on the $11'600$-points test dataset is $1.34\cdot10^{-2}$. 
Fig. \ref{fig:DeltaAngle_OnlyAngle} depicts the comparison between the predicted and the actual trajectory of the angle $\delta(t)$ and the frequency $\omega(t)$. The best and worst $\{\delta, \omega\}$ estimation during different active power inputs $P_1$ in terms of $L_2$ error on both training and test sets are depicted in the left and right side of the figure, respectively. To extract the frequency $\omega$ we differentiate the signal associated with the angle $\delta$. To this end, we numerically approximate the derivative of a function as:
  $  \omega(t)=\lim_{h \rightarrow0} \tfrac{\delta(t+h)-\delta(t)}{h}$.
The value of $h$ depends on the simulation time step. In this study, we generated the trajectories with a fixed step of $h=0.1s$.
In future work, we will use automatic differentiation \cite{baydin2018automatic} to extract the frequency directly from the physics-informed neural network. 
It can be observed that the physics-informed neural network is able to predict the trajectory of the angle $\delta(t)$ with high accuracy, and that the frequency signal $\omega(t)$ can be successfully recovered using numerical differentiation.


After training, we evaluate the neural network performance in terms of computational speed required for solving the differential equation defined by \eqref{Eq:Swing_SMIB}. For 100 different initial conditions of \eqref{Eq:Swing_SMIB}, the $ode45$ solver takes on average 0.45~\,s to solve the differential equations and the neural network only 0.016\,s, resulting to a speed-up of factor of 28. We expect that for larger systems the computational speed-up will be even higher, as solving large-scale differential equations is computationally very expensive, whereas the evaluation of a trained neural network remains computationally low even for large network sizes. Additionally, and most importantly, the physics-informed neural network can directly determine $\delta$ at any specified time step $\delta(t_1,P_1)$, whereas numerical methods always have to start integrating from the boundary conditions at $t=t_0$ until they reach $t=t_1$. The computational time for evaluating any random time step (e.g. at $t_1=10$\, s) is $4\cdot10^{-3}$\,s, whereas integrating from $t_0=0s$ up to $t_1=10$\, s with the $ode45$ solver takes 0.35~s, resulting to a \emph{substantial speed-up of almost two order of magnitude} for the physics-informed neural network (87 times to be exact). This illustrates the capability of physics-informed neural networks to predict directly the solution to higher-order differential equations with high accuracy and low computational cost, offering significant advantages over classical numerical integration tools. 

\subsubsection{Predicting both angle $\delta$ and frequency $\omega$ as separate neural network outputs}
Within our investigations, we also attempted to train a physics-informed neural network that considers $\delta$ and $\omega$ as separate outputs, essentially setting $f_{\omega} = \dot{\delta}-\omega$ and $f_{\delta} = m_1 \dot{\omega} + d_1 \omega + B_{12}V_1V_2\sin(\delta) - P_1$. To obtain the lowest $L_2$ error in this case, we had to select again a set of $N_u$ = 40 randomly distributed initial and boundary data, a 5-layer neural network with 10 neurons per hidden layer, but a set \mbox{$N_f$ = $50'000$}  collocation point (instead of $8'000$ in the previous case) . The model training took approximately 30 minutes as more collocation points $N_f$ are required to obtain a satisfactory prediction error. Considering that $\delta(t)$ and $\omega(t)$ are predicted as separate outputs, the relative $L_2$ errors between the exact and predicted solutions are $9.43\cdot10^{-2}$ and $1.51 \cdot 10^{-1}$, respectively, and are higher than for the $NN_{\delta}$ structure. It becomes obvious that the neural network architecture with the single output $\delta$ (and subsequent numerical differentiation to determine $\omega$) is preferable in terms of training time and predictive accuracy.  


\subsection{Data-driven discovery of inertia and damping coefficients through physics-informed neural networks}
In this subsection, we evaluate the performance of the physics-informed neural network to predict system inertia and damping from observed trajectories. In this case study, we assume that $m_1$ and $d_1$ are unknown, and instead we have a set of limited training datapoints $\{t, P_1, \delta\}$. Contrary to the usual practice of first training a neural network and then using it, our objective here is exploit the physics-informed neural network training procedure to determine $m_1$ and $d_1$. To illustrate the effectiveness of this approach, we perform this analysis for 10 different pairs of $\{m_1, d_1\}$ and evaluate the average predictive accuracy.
We select a set of \mbox{$N_u$ = 100} randomly distributed points across the spatio-temporal domain from the exact solutions of \eqref{Eq:Swing_SMIB} for each inertia level. 
A 5-layer neural network with 30 neurons per hidden layer is trained for each inertia level with the corresponding trajectories in order to predict the system parameters and $\delta(t)$. 
The resulting average errors for predicting $m_1$ and $d_1$ over the 10 different cases are 0.74\% and 1.28\%, respectively. 
The average training time of the neural network to identify the system parameters was less than 60 seconds. This means that with a limited training dataset, and within 60 seconds, we can accurately predict the inertia and damping level of a system. 
Considering that the swing equation \eqref{Eq:Swing_SMIB} is often used to approximate the aggregate dynamic behavior of large power systems, these results demonstrate that physics-informed neural network show substantial potential to not only accurately derive $\delta$ and $\omega$ but also predict both system inertia and damping. Last but not least, the relative $L_2$ errors between the exact and predicted solutions for the phase angle are less than $10^{-1}$ over the 10 different cases of $\{m_1, d_1\}$. This shows the potential of physics informed neural networks to be used as a dynamic state estimator, when the model parameters are unknown \cite{Zhao2019PSDSE}.



\section{Discussion and Outlook}
This work introduces for the first time in power systems a neural network training procedure that explicitly considers the underlying differential and algebraic equations describing power system behavior. This unlocks a series of opportunities in power systems, as physics-informed neural networks may be able to accurately determine the solution of differential-algebraic sets of equations several orders of magnitude faster than traditional methods relying on numerical integration. Still, to unlock this potential, there are several challenges to be addressed.  
\paragraph{Number of training data}
Besides the limited number of training data, physics-informed neural networks as described in this paper need to generate a substantial number of collocation points. In our case studies, we used $N_u=40$ points as input data and $N_f=8'000$ collocation points. It is expected that for larger systems, a much larger number of collocation points will be necessary, which will result to a longer training time. In our future work, we plan to investigate methods using Runge-Kutte integration schemes such as the ones proposed in \cite{RAISSI2019686} which can eliminate the need for collocation points. 
\paragraph{Scalability}
Although the swing equation is a good first approximation for first-swing instability, and single-machine infinite-bus systems are still used as aggregate models of large power systems, we still need to explore what are the computational needs if we were to apply these methods in large scale power systems and how to address the associated challenges related to the neural network training. Particularly, the comparison with numerical solvers and approximation techniques like polynomial fits will serve as a benchmark.
\paragraph{Range of applications}
As shown in this paper, physics-informed neural networks can determine two orders of magnitude faster the rotor angle and frequency at any time instant for uncertain power inputs. At the same time, they can accurately identify uncertain parameters such as inertia and damping. Future applications must also assess cases that include both stable and unstable equilibria, a wide range of different dynamic phenomena, including small-signal stability, voltage stability and converter dynamics \cite{Misyris2019}, discrete events, such as protection actions, as well as power system optimization, among numerous others. In our simulation study, we observed high accuracy for a single stable swing prediction, but for different regimes such as multiple oscillations or unstable conditions, different physics-informed neural networks might have to be trained. We also need to examine if such neural networks can capture discrete events, such as protection actions, or if we need to develop a hybrid approach, using physics-informed neural networks as a numerical solver only during the continuous dynamics before and after a discrete event. For power system applications, physics-informed neural networks can (and should) be combined with neural network verification methods, see~\cite{venzke2019verification}. In this way, they would no longer be considered a black box, but instead we would be able to extract formal guarantees for their behavior.

\section{Conclusions}
To the best of our knowledge, for power system applications, this is the first paper to propose physics-informed neural networks. Explicitly considering the power system governing equations, we are able to determine the solution of differential-algebraic systems of equations at a fraction of the time required for conventional numerical approaches. Physics-informed neural networks require substantially less training data, while achieving high accuracy, due to the inclusion of the underlying swing equation. This paper introduces the general framework and presents results for a single-machine infinite-bus system. In our case studies, we demonstrate how physics-informed neural networks can accurately determine the rotor angle and frequency \emph{87 times} faster than conventional numerical methods. We further demonstrate their successful identification of uncertain system parameters such as inertia and damping from a limited set of input data. Our results showcase the potential for successful application of these methods in larger systems, unlocking a series of opportunities for power system security and optimization, achieving good accuracy and high computational speed. Future work will explore a series of possible applications and potential improvements in the training procedure. 
\section*{Acknowledgement}
\vspace{-3pt}
{\footnotesize This work is supported by the multiDC project funded by Innovation Fund Denmark, Grant No. 6154-00020B.}
\vspace{-3pt}

\bibliographystyle{IEEEtran}
\bibliography{References}

\end{document}